\documentclass{elsarticle}

\usepackage{hyperref,mathtools}
\newcommand{\bfU}{\mathbf{U}}
\newcommand{\bfB}{\mathbf{B}}
\newcommand{\bfS}{\mathbf{S}}
\newcommand{\bfF}{\mathbf{F}}
\newcommand{\bfx}{\mathbf{x}}
\newcommand{\bfv}{\mathbf{v}}
\newcommand{\bfe}{\mathbf{e}}
\newcommand{\bfV}{\mathbf{V}}
\newcommand{\bfI}{\mathbf{I}}
\begin{document}

\begin{frontmatter}

\title{{\tt MPI-AMRVAC}: a parallel, grid-adaptive PDE toolkit}

%\author{Rony Keppens\fnref{myfootnote}}

\author{Rony Keppens\fnref{mycorrespondingauthor}}
\fntext[mycorrespondingauthor]{Corresponding author}
\address{Centre for mathematical Plasma Astrophysics, KU Leuven, Belgium}
\ead{rony.keppens@kuleuven.be}
\ead[url]{perswww.kuleuven.be/Rony\_Keppens}
\author{Jannis Teunissen}
\address{CWI, Amsterdam, The Netherlands}
\author{Chun Xia}
\address{Yunnan University, School of Physics and Astronomy, Kunming, Yunnan, PR China}
\author{Oliver Porth}
\address{UvA, Amsterdam, The Netherlands}
\begin{abstract}
We report on the latest additions to our open-source, block-grid adaptive framework {\tt MPI-AMRVAC}, which is a general toolkit for especially hyperbolic/parabolic partial differential equations (PDEs). Applications traditionally focused on shock-dominated, magnetized plasma dynamics described by either Newtonian or special relativistic (magneto)hydrodynamics, but its versatile design easily extends to different PDE systems. Here, we demonstrate applications covering any-dimensional scalar to system PDEs, with e.g. Korteweg-de Vries solutions generalizing early findings on soliton behaviour, shallow water applications in round or square pools, hydrodynamic convergence tests as well as challenging computational fluid and plasma dynamics applications. The recent addition of a parallel multigrid solver opens up new avenues where also elliptic constraints or stiff source terms play a central role. This is illustrated here by solving several multi-dimensional reaction-diffusion-type equations. We document the minimal requirements for adding a new physics module governed by any nonlinear PDE system, such that it can directly benefit from the code flexibility in combining various temporal and spatial discretisation schemes. Distributed through {\tt GitHub}, {\tt MPI-AMRVAC} can be used to perform 1D, 1.5D, 2D, 2.5D or 3D simulations in Cartesian, cylindrical or spherical coordinate systems, using parallel domain-decomposition, or exploiting fully dynamic block quadtree-octree grids.
\end{abstract}


\end{frontmatter}

%\linenumbers

\section{Introduction}
In contemporary astrophysical research, numerical modeling forms a vital ingredient, almost invariably handling strongly nonlinear flows and plasma dynamics (i.e., the fourth and most abundant state of known matter in our universe). Many open source codes~\cite[e.g][]{pluto2007,pluto2012,flash2000,nirvana1999,nirvana2008,athena2008,ramses2002,mpiamrvac2018} are actively developed and used, which focus on shock-dominated scenarios in gases or plasmas, enriched by radiative processes, gravitational interactions, as well as various (energy) transport and exchange mechanisms, where the equations of (magneto)hydrodynamics or (M)HD form the core application. These (M)HD equations, covered in various textbooks~\citep[e.g.][]{Hansbook3}, return in many aerodynamical or engineering scenarios. This continuously drives the need for advanced numerical techniques to handle (also transsonic and supersonic) flows about obstacles (airplanes or satellite re-entry problems), ventilation flows through ducts, or the generic behavior of electrically conducting fluids.

Many astrophysical applications must handle a vast range of spatial scales, so it is costumary to incorporate adaptivity in the numerical solution, where different strategies exist: dynamically relocating a fixed number of grid points ($r$-refinement, see e.g.~\cite{book2001}); using a dynamic means to increase or decrease the number of grid cells by varying the cell sizes ($h$-refinement); or ensuring that the local polynomial representation of the solution throughout a grid cell employs a differing order ($p$-refinement). Various open source codes~\cite[e.g.][]{pluto2012,flash2000,nirvana1999,ramses2002,mpiamrvac2012,Collins2010} exploit $h$-refinement, where the mesh has various levels of successively finer grids, organized in a hierarchical manner. We will specify the further discussion to {\tt MPI-AMRVAC}\footnote{\url{http://amrvac.org}}~\cite{mg2019,mpiamrvac2018,mpiamrvac2014,mpiamrvac2012}, which evolved from a patch-based adaptive mesh refinement (AMR) framework~\cite{Keppens2003}, to a purely block-octree AMR implementation~\cite{mpiamrvac2012,mpiamrvac2018}. 

\subsection{{\tt MPI-AMRVAC} code basics}
{\tt MPI-AMRVAC} has been in continued development for more than a decade~\cite{Keppens2003}, with modern applications ranging from magnetospheric physics at Earth~\cite{Leroy2017} or in the Jovian system~\cite{Chane2017}, over solar physics challenges~\cite{Ruan2019,Zhou2018,Xia2017}, to more exotic astrophysical processes such as those encountered in supergiant X-ray binaries~\cite{Ileyk2019}. The code has heritage to the original {\tt Versatile Advection Code} (hence {\tt VAC})~\cite{vac1996,Toth1996,Keppens1999}, which solved (near) conservative sets of mainly hyperbolic partial differential equations (PDEs), and specialized in MHD problems~\citep[e.g.][]{keppenstoth1999,KeppensKH1999,tothdivb2000}.  In this paper, we demonstrate that {\tt MPI-AMRVAC} is well suited to handle fairly diverse systems of PDEs, that may even deviate from being advection-dominated problems. Indeed, the modular design makes it easy to introduce PDE systems of the form
\begin{equation}
\partial_t \bfU + \nabla \cdot \bfF(\bfU) = \bfS(\bfU,\partial_x\bfU,\partial^2_{xx}\bfU, \ldots,\bfx,t) \,, \label{eq}
\end{equation} 
where the set of variables $\bfU$ is subject to fluxes $\bfF$ and source terms $\bfS$, where all variables $\bfU=(U_1, U_2, \ldots, U_{m})$ are to be solved for their spatiotemporal $U_i(\bfx, t)$ behavior. The spatial coordinates $\bfx$ may be 1D, 2D or 3D Cartesian coordinates, or could be polar, cylindrical or spherical in nature. 

The HD system in particular has (conservative) variables $\bfU=(\rho, \rho\bfv, E)^T$ consisting of density $\rho$, momentum density vector $\rho\bfv$ (with velocity $\bfv$) and total energy density $E$ (combining kinetic with internal energy in $\rho v^2/2+p/(\gamma-1)$ with pressure $p$ and parameter $\gamma>1$). In the hyperbolic PDE system for HD, the fluxes $\bfF$ typically split up into an advection $\bfF_{\mathrm{ad}}\equiv \bfv\bfU$ and a non-advective flux $\bfF_{\mathrm{na}}$, and speed magnitudes $v\equiv | \bfv| $ may be below, equal or above the local physical sound speed. In MHD, also the magnetic field vector $\bfB$ enters as a variable. To handle discontinuous, shocked flow problems in (M)HD, it is imperative to use conservative numerical schemes~\cite[e.g.][]{toro97,Leveque,Hansbook3}, which usually handle fluxes $\bfF$ in a manner exploiting the (approximate) solution of local Riemann problems (i.e., initial conditions seperating two constant states $\bfU_{\mathrm{l}}$ and $\bfU_{\mathrm{r}}$, to the left and right of the discontinuity, respectively). For both the HD and the MHD equation set, in their Newtonian as well as special relativistic variant~\cite{Bart2008,mpiamrvac2012}, codes like {\tt MPI-AMRVAC} offer a wide variety of spatio-temporal discretizations, to advance initial conditions augmented with boundary prescriptions, in 1D, 2D or 3D configurations. For {\tt MPI-AMRVAC}, the {\tt Fortran} source code is documented at {\tt amrvac.org}, and available on {\tt GitHub}. Making use of {\tt Doxygen}\footnote{\url{http://www.doxygen.nl}}, the inline documentation is automatically turned into dependency graphs, flow charts, and searchable source code, which is updated daily to reflect the current status of the code development.

\subsection{Adding a new PDE system}

Any system of the form given by Eq.~(\ref{eq}) may be added to the framework, whose source code is typically located in {\tt amrvac\//src} (and an environment variable {\tt AMRVAC\_DIR} is to be set to locate this {\tt amrvac} directory). The minimal requirement for adding a new system is to create a corresponding system module (a subdirectory {\tt amrvac\//src\//newsystem}) quantifying the variables $\bfU$, fluxes $\bfF$ and source terms $\bfS$. In the generic physics module {\tt amrvac\//src\//physics\//mod\_physics.t}, procedure pointers are initialized and their calling interface is predefined, for use in the PDE systems to implement. Among other procedures, this generic module contains {\tt phys\_get\_flux} and {\tt phys\_add\_source} interfaces, which must  be fully provided in the module {\tt amrvac\//src\//newsystem\//mod\_newsystem\_phys.t} of a newly added system. For example, the available (M)HD systems are found in {\tt amrvac\//src\//hd} and  {\tt amrvac\//src\//mhd}, where the system-specific initializations are handled by an {\tt mod\_hd.t} and {\tt mod\_mhd.t} activation module, while the actual equations (variable definitions, fluxes, sources and corresponding time step constraints) are to be found in {\tt mod\_hd\_phys.t} and {\tt mod\_mhd\_phys.t}. Besides the mentioned interfaces for providing fluxes and sources, other procedures of interest are 
\begin{itemize}
\item {\tt phys\_add\_source\_geom} for the handling of geometric source terms, needed when solving the same system in polar, cylindrical or spherical coordinates;
\item {\tt phys\_get\_v\_idim} to specify an advection velocity in the {\tt idim} direction, when an advective flux $\bfF_{\mathrm{ad}}$ is to be used. Note that this direction refers to $x$, $y$ or $z$ in Cartesian cases, while it is e.g. $r$, $\varphi$ in a polar grid;
\item {\tt phys\_get\_dt} to provide a system-specific time step constraint, that would be in addition to the usual Courant-Friedrichs-Lewy (CFL) limit for explicit time stepping schemes;
\item {\tt phys\_get\_cmax} and {\tt phys\_get\_cbounds} to quantify the maximal physical propagation speed and any minimal and maximal bound on that speed, useful for computing the CFL timestep limit, or in use for the simplest of any approximate Riemann solver methods (i.e. the local Lax-Friedrich or TVDLF method~\cite{rusanov,Toth1996}, as well as its HLL extension~\cite{hll}, which are used heavily in (M)HD applications). 
\end{itemize}
If the system of equations differentiates between conservative $\bfU$ and primitive variables $\bfV$, like the set $\bfV=(\rho,\bfv,p)^T$ with pressure $p$ for the set of $\bfU=(\rho,\rho\bfv,E)$ in HD, one can provide conversion formulae from conservative to primitive in the procedure {\tt phys\_to\_primitive}, while its reverse {\tt phys\_to\_conserved} switches primitive variables back to conservative ones.  To benefit optimally from the dimension-independent implementation of our code, these routines best exploit the {\tt LASY} syntax~\cite{Lasy1997}, which means writing the coordinate array $\bfx$ as 
\begin{verbatim}
x(ixI^S,1:^ND) 
\end{verbatim}
which will expand the segment \verb|^S| when the dimensionality \verb|^ND=2| to 
\begin{verbatim}
x(ixImin1:ixImax1,ixImin2:ixImax2,1:2) 
\end{verbatim}

\subsection{Adaptive mesh refinement and parallelization}

When implementing a new system in {\tt MPI-AMRVAC} according to the procedure just explained, one can directly benefit from its dimension-independent way to solve the system~(\ref{eq}) with a variety of time stepping schemes, splitting strategies for handling sources, and exploit its parallel implementation to run efficiently on laptops to the most modern supercomputers. The {\tt Message Passing Interface} or {\tt MPI} based parallelization can always exploit a domain-decomposition by specifying block sizes that equally divide up the computational domain (e.g., one may decide to use a $200\times 200$ 2D mesh, divided into 400 blocks of size $10\times 10$). The {\tt AMR} stands for Adaptive Mesh Refinement, where one relies on a block-based quadtree-octree (in 2D-3D) means of hierarchically adjusting the computational mesh to the evolving solution. The blocksize can be specified and adjusted by the user. A fair variety of automated as well as user-specific means to set (de)refine criteria is available, and they work for all dimensionalities and coordinate systems provided. The excellent scaling of the parallel implementation has been demonstrated in previous work~\cite{mpiamrvac2014}, while the specification of a new system of equations is essentially devoid of any {\tt MPI} procedures, except for trivial \verb|mpistop(``error_message'')| interfaces to {\tt MPI\_ABORT} calls for catching erronous input parameter specifications. 

The {\tt AMR} strategy operates as follows. The user specifies a block size {\tt block\_nx1, block\_nx2, \ldots} for each dimension of the problem at hand, and a conforming domain size in number of grid cells {\tt domain\_nx1, domain\_nx2, \ldots} at the lowest resolution. The actual domain physical extent is specified by coordinate pairs, such as {\tt xprobmin1, xprobmax1} for the minimal and maximal $x$-coordinate, respectively. A maximal number of refinement levels is set through {\tt refine\_max\_level}, and actual adaptive runs imply that this maximal refinement level is $\geq 2$, while a unit value realizes a pure domain-decomposition computation. When adaptivity is turned on, all blocks at grid levels below {\tt refine\_max\_level} evaluate a user-selected refinement criterion in every grid cell. The default refinement criterion is a Lohner type estimator~\cite{lohner}, where we essentially quantify local weighted second derivates (for details, see~\cite{mpiamrvac2012}), and this for a user-selected set of variables. This then provides an error in all grid cells of the evaluated blocks. A block is then refined using a fixed refinement ratio of 2 (in 2D this implies splitting the block in 4, in 3D each block creates 8 children blocks) as soon as it has a single point whose error exceeds a user-set tolerance. For a block to be coarsened, all its cells must have the error below a user-set fraction of the previous tolerance. Our implementation also allows the user to intervene with the automated refinement, either overruling or enforcing refinement when necessary.

In this paper, some exemplary problems are presented, where our flexible means of post-processing multi-dimensional data is exploited. The latter comprises possibilities to convert data files on the fly (or after the computation is completed) to data formats directly importable in open-source visualization software like {\tt Paraview}\footnote{\url{https://www.paraview.org}} or {\tt VisIt}\footnote{\url{https://wci.llnl.gov/simulation/computer-codes/visit}}. Alternatively, one may use the many provided {\tt python} scripts to e.g. regrid the hierarchically meshed data to a uniform coverage, and use free plotting packages. 

\subsection{Multigrid functionality}
The idealized (M)HD systems which feature in many astrophysical applications are hyperbolic in nature, but when effects like thermal conduction, viscosity, or resistivity are incorporated, parabolic source terms appear. Typical diffusion terms may well render the standard explicit time stepping strategies impractical, as then the time step constraint scales with $\Delta t \propto h^2$, for grid spacing $h$, prohibiting the use of ultra-high resolution. This is in direct conflict with the usual advantage offered by an AMR code, allowing for extreme resolutions at affordable costs. To alleviate this drawback, a recent extension of our code is its coupling to a fast elliptic solver~\cite{mg2019}. The block-adaptive grid used in {\tt MPI-AMRVAC} suggests the use of a (geometric) multigrid strategy, where Poisson or Helmholtz equations, with variable coefficients, can be solved in a highly scalable fashion. This recent addition is e.g. useful for handling the Maxwell equation $\nabla\cdot\bfB=0$ in multi-dimensional MHD problems, but may also be used to implement particular implicit-explicit (IMEX) discretizations of PDEs containing stiff sources (as demonstrated further on), to handle any typical parabolic terms, to implement incompressible (M)HD equations, or to solve astrophysical applications involving self-gravity. In the next section, we include cases where the multigrid functionality proves helpful to efficiently compute PDE solutions.

\section{Example applications}
\subsection{Korteweg-de Vries computations}
As a first demonstration, we present 1D and 2D solutions for a nonlinear scalar equation known as the Korteweg-de Vries equation. This equation combines nonlinear advection with a source term containing a third-order derivate,
\begin{equation}
\partial_t \rho + \nabla\cdot (\frac{1}{2}\rho^2\bfe) = - \delta^2\sum_{i=1}^D \partial_{xxx_i} \rho  \,, \label{kdv}
\end{equation}
where $\bfe=\sum_{i=1}^D \hat{\bfe}_i$ in a $D$-dimensional setup, with Cartesian unit vectors $\hat{\bfe}_i$. In 1D, this recovers the original Korteweg-de Vries or KdV equation 
\begin{equation}
\partial_t\rho+\rho\partial_x\rho+\delta^2\partial_{xxx} \rho=0 \,, \label{kdv1d}
\end{equation}
where $\delta$ is a fixed parameter. For $\delta=0$, we get the nonlinear Burgers equation, which can be used to test shock formation through wave steepening and its numerical realization~\cite{keppensporth14}. In~\cite{Zabusky1965}, the 1D KdV equation~(\ref{kdv1d}) was solved numerically on a periodic domain $x\in[0,2]$, with $\delta=0.022$, and initial condition provided by $\rho(x,t=0)=\cos(\pi x)$. This classic paper~\cite{Zabusky1965} documented how the numerical solution showed the spontaneous development and interaction of multiple solitons, where the nonlinear term causing wave steepening is balanced by the dispersive source term to maintain their integrity. 

\begin{figure}[t]
\centerline{\includegraphics[width=\textwidth]{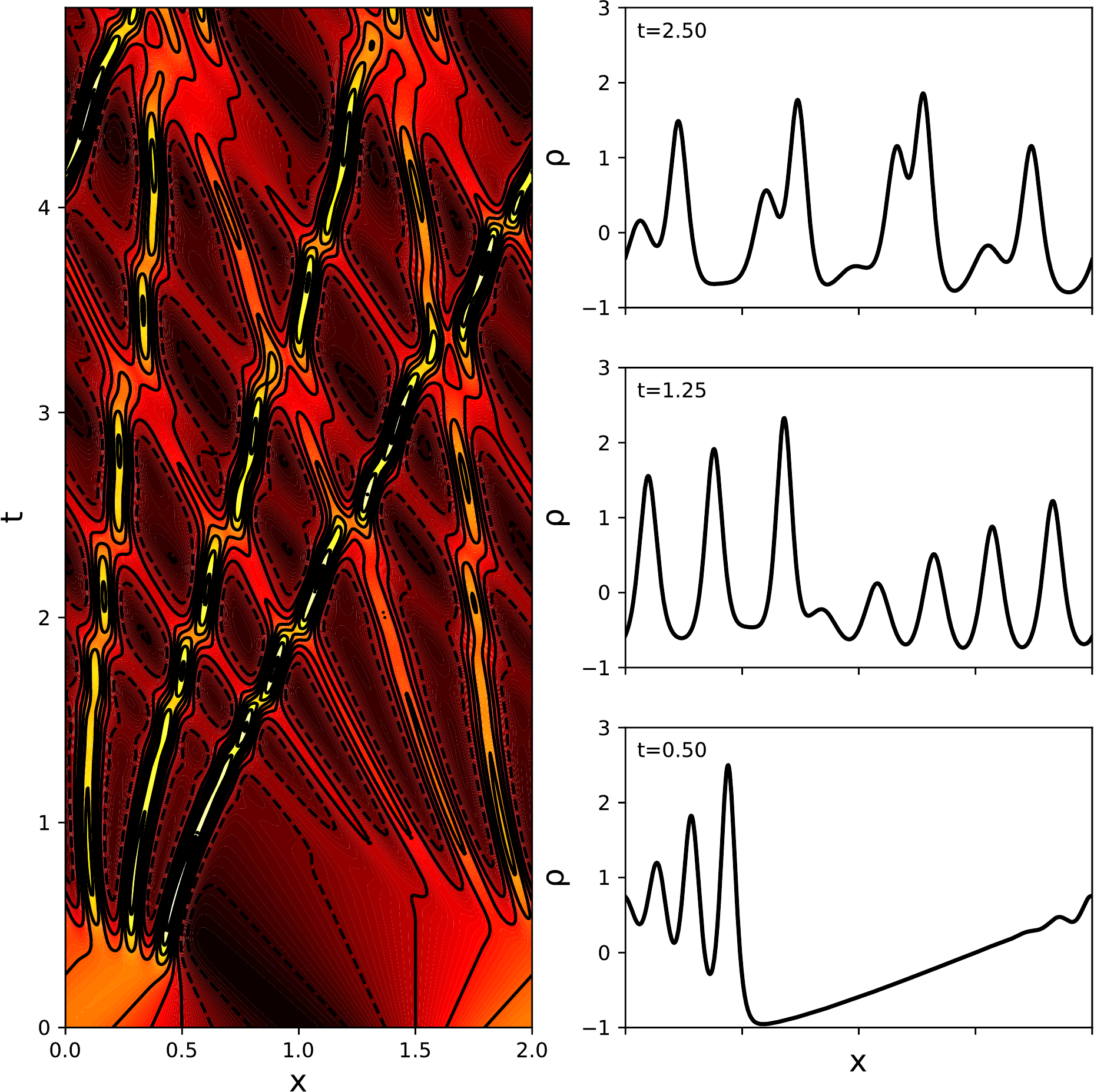}}
\caption{A 1D KdV simulation, showing $\rho(x,t)$ at left in a contour view, with selected $\rho(x)$ variations obtained at the times indicated.}\label{f-kdv1}
\end{figure}

In {\tt MPI-AMRVAC}, the scalar Eq.~(\ref{kdv}) is implemented in a {\tt amrvac\//src\//nonlinear} module, and a Boolean {\tt kdv\_source\_term} can activate the addition of the dispersive source term. This source term can be evaluated using a fourth order central difference evaluation, requiring three ghost cells to each block when a domain-decomposition strategy is used. We solved the KdV equation on a time interval $t\in[0,5]$, using 600 grid points, subdivided into 10-cell blocks. This test was run without AMR. A Courant number of 0.9 is used for the CFL condition, where the local absolute value of the scalar $\rho$ sets the maximal signal speed. An additional time step limit is enforcing $\Delta t \leq 0.9 (\Delta x)^3/(3\sqrt{3}\delta^2/2)$, where we follow a prescription specified by~\cite{Lee2017}. For the handling of fluxes, we use a conservative finite difference scheme using a fifth-order, monotonicity preserving MP5 reconstruction~\cite[for details, see][and references therein]{mpiamrvac2014}. The combination of the conservative finite difference scheme, a three-step Runge-Kutta time integrator, and the central difference source evaluation, makes that the numerical solution conserves $\int_0^2 \rho \, dx$ exactly, which is a known property of the KdV equation. In Fig.~\ref{f-kdv1}, we show the solution $\rho(x,t)$ in a contour plot view on the left, where one recognizes how the original cosine variation leads to three sharply peaked solitons that eventually travel forward through the domain, while five weaker backward propagating solitons emerge somewhat later. Their repeated interactions as they pass periodically across the boundaries are very well represented. At right in Fig.~\ref{f-kdv1}, selected instantaneous $\rho(x)$ profiles at times $t=0.5$, 1.25 and 2.5 are provided. 

\begin{figure}[t]
\centerline{\includegraphics[width=\textwidth]{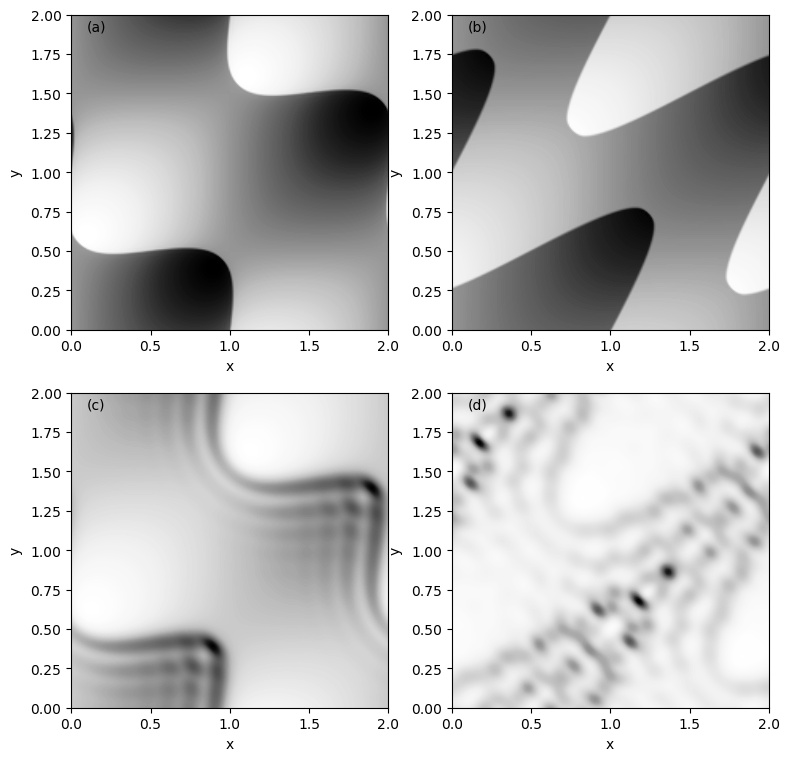}}
\caption{Instantaneous solutions at $t=0.4$ (left panels (a) and (c)) and $t=1$ (right panels (b) and (d)), comparing Burgers (top (a) and (b)) with KdV (bottom (c) and (d)).}\label{f-kdv2}
\end{figure}

As stated, any multi-dimensional variant of Eq.~(\ref{kdv}) can now easily be simulated as well, and we compare 2D solutions for a Burgers equation (where $\delta=0$) to the 2D version of the previous KdV test ($\delta=0.022$). The double periodic domain $[0,2]\times[0.2]$ is initialized with $\rho(x,y)=\cos(\pi x)\sin(\pi y)$, using a $200\times 200$ grid in blocks of $10\times 10$, again in domain-decomposition mode without AMR for simplicity, since the computational cost for solving a single scalar equation is rather small. We use the same scheme combination as before, only reducing the source-related time step limit to $\Delta t \leq 0.4 (\Delta x)^3/(3\sqrt{3}\delta^2/2)$. In Fig.~\ref{f-kdv2}, snapshots of the density profile at $t=0.4$ (left column) and $t=1$ (right column) are compared for the Burgers equation (top row) versus the KdV system (bottom row). Note the clear shock-dominated solution for the 2D Burgers variant, while the KdV equation again shows soliton-like features developing spontaneously. The patterns observed in the KdV solutions remind us of ripples in (shallow) water and their interactions. This is not surprising, since the original KdV equation arose from analyzing a specific limit of the shallow water equations, to which we turn attention next.

\subsection{Shallow water test problems}

The shallow water equations can be formulated as
\begin{equation}
\partial_t \begin{pmatrix} 
h \\ 
h\bfv 
\end{pmatrix}
+ \nabla\cdot 
\begin{pmatrix}
\bfv h \\ 
\bfv\bfv h + \frac{1}{2}h^2 \bfI
\end{pmatrix} = \bfS \,, \label{sw}
\end{equation}
where one solves for the height profile $h(x,y,t)$ with (height-averaged) speeds $\bfv=\left(v_x(x,y,t), v_y(x,y,t)\right)$ affected by gravity. $\bfI$ is the unit tensor. The above formulation exploits a dimensionless form of the shallow water equations, where lengths are scaled as $\bar{h}=h/a$ and times $\bar{t}=t\sqrt{g/a}$, where $a$ is a reference length unit (e.g. $a$ can be set to 1 meter) and Earth's gravitational acceleration is $g=9.8 \, \mathrm{m}/\mathrm{s}^2$. Eq.~(\ref{sw}) is to be read with $h\rightarrow \bar{h}$ and similarly for all quantities, while the source term $\bfS$ may introduce resistance to the flow and bottom topology~\cite{Delis2005,Zoppou2000}. 
The shallow water equations make sense in 2D, and they can be solved on either a Cartesian domain, or a polar $(r,\varphi)$ grid. The latter introduces geometric source terms, in particular a source $(\frac{1}{2}h^2+hv_\varphi^2)/r$ for the radial momentum $hv_r$, as well as a term $-h v_r v_\varphi/r$ for the azimuthal component $hv_\varphi$. These geometric source terms are then in addition to possible resistance or flow bed topology terms encoded in $\bfS$. 

In {\tt MPI-AMRVAC}, the above equation set is in fact available within the HD module, where one recognizes the fact that system~(\ref{sw}) is identical to the subset of mass and momentum conservation laws in the Euler equations, with the `pressure' given by $h^2/2$. Hence, a switch to avoid using an energy variable $E$ is introduced in our {\tt amrvac\//src\//hd} system, in which case the `pressure' $p=c_{\mathrm{ad}}\rho^\gamma$, where a polytropic relation between density $\rho$ and pressure introduces two free parameters, $c_{\mathrm{ad}}$ and $\gamma$. The `sound' signal speed (squared) is then $c^2=\gamma c_{\mathrm{ad}} \rho^{\gamma-1}$, and the shallow water system arises for the identification $\rho\equiv h$, $c_{\mathrm{ad}} =0.5$ and $\gamma=2$. In fact, our HD system module has various switches for handling either subcases of the full Euler system (or Navier-Stokes when activating viscosity), or extensions of the Euler system where additional `dust' species are handled as pressureless fluids, coupled to the Euler gas by means of drag terms~\cite{mpiamrvac2014,hendrix2016}. Hence, a particular equation module can serve multiple purposes.

\begin{figure}
\centerline{\includegraphics[width=\textwidth]{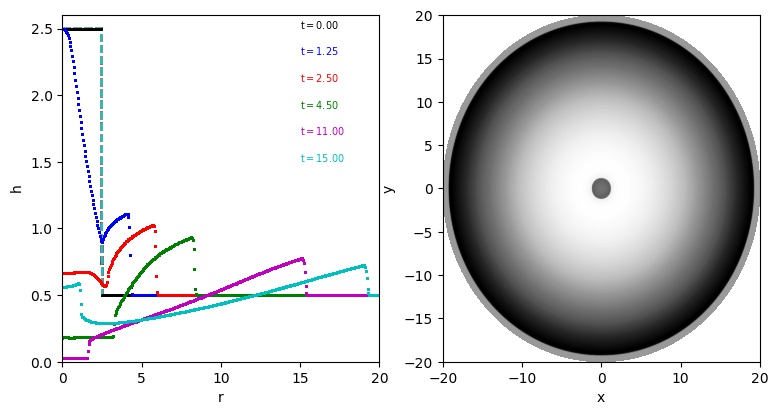}}
\caption{A circular dam break problem, solved on a polar adaptive grid. Left: scatter plots of $h(r)$ at selected times. Right: contour plot of $h(r,\varphi)$ at $t=15$.}\label{f-swpolar}
\end{figure}

As a stringent test of the shallow water implementation, we target the reference test introduced by~\cite{toro01} and discussed in~\cite{Delis2005}, mimicking a circular dam break where there are no extra source terms $\bfS$. The test is in fact a 1D Riemann problem in a cylindrical configuration, having water height $h_{\mathrm{in}}=2.5\, \mathrm{m}$ inside a circular dam of radius $r_{\mathrm{dam}}= 2.5 \,\mathrm{m}$, while the exterior has a water height of $h_{\mathrm{ext}}=0.5\, \mathrm{m}$. We choose to solve this problem twice, once on a 2D polar grid on a disk of radius 20 meters, and once in a 2D Cartesian domain of size $40 \,\mathrm{m} \times 40 \,\mathrm{m}$. The solutions must obviously agree, but the latter one could suffer from artificial deformations when solving an azimuthally symmetric problem on a square grid. In this problem, we use AMR and exploit a total of 3 refinement levels in both the polar and the Cartesian setup.

In Fig.~\ref{f-swpolar}, we show at right the solution at dimensionless time $t=15$ (physical time $t=4.79 \,\mathrm{s}$), as a contour plot of the $h(r,\varphi)$ solution. At left, the same solution, as well as several earlier snapshots (between $t=0$ and $t=15$) are plotted as a scatter plot where all gridcells are visualized. The symmetry (i.e. the 1D nature of this problem) is perfectly preserved, as each dot in the scatter plot is simply repeated for as many azimuthal grid cells as used. We actually adopted a base grid level of $100\times 100$ in $(r,\varphi)$, augmented with two additional refinement levels, triggered on height and momentum variations, effectively showing a $400\times 400$ resolution. Block sizes of $10\times 10$ are used, and the boundary conditions use a $\pi$-periodic treatment across the $r=0$ pole~\cite{Bart2007}, periodicity in $\varphi$, and a zero gradient (Neumann) extrapolation at $r=20$. As spatio-temporal integration method, we combined an HLL scheme~\cite{hll}, a threestep Runge-Kutta, and a Koren limiter~\cite{Koren1993} based reconstruction procedure, with a Courant number of 0.9.  

\begin{figure}
\centerline{\includegraphics[width=\textwidth]{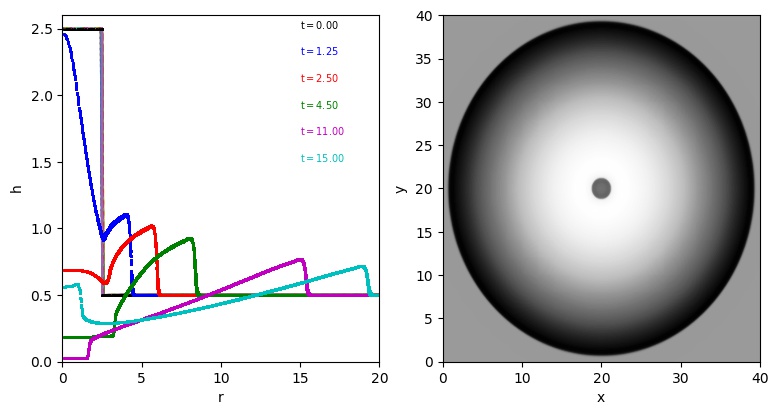}}
\caption{The same as Fig.~\ref{f-swpolar}, but now solved on an adaptive Cartesian grid.}\label{f-sw}
\end{figure}

The same problem, now solved on a Cartesian grid, under otherwise identical settings (except for adjusting the domain and the boundary condition on the $y$-borders), is displayed in the same fashion in Fig.~\ref{f-sw}. Note how the contour plot at $t=15$ is visually indistinguishable from its polar variant, while the Cartesian grid now obviously samples the radial profile more frequently when plotted as a scatter plot of $h(r)$ (left panel). For both the polar and Cartesian realization, the AMR that originally locates at the initial discontinuity essentially spreads across the full domain, capturing the outward propagating shock front and the rapid central height variation within $r\leq 5$ as seen in the left panels of both figures.

\subsection{Hydrodynamical tests}

In this section, we show two example tests for the HD system, where now the equation for the energy density variable $E=\rho v^2/2+p/(\gamma-1)$ is included. We set $\gamma=5/3$, the standard value for an ideal monoatomic gas. In the first test, a convergence study of an Euler solution is made on uniform grids, while the second test shows solutions for the compressible Navier-Stokes system, using AMR in combination with embedded boundaries.

\subsubsection{Gresho-Chan vortex} 
\begin{figure}
\centerline{\includegraphics[width=0.49\textwidth]{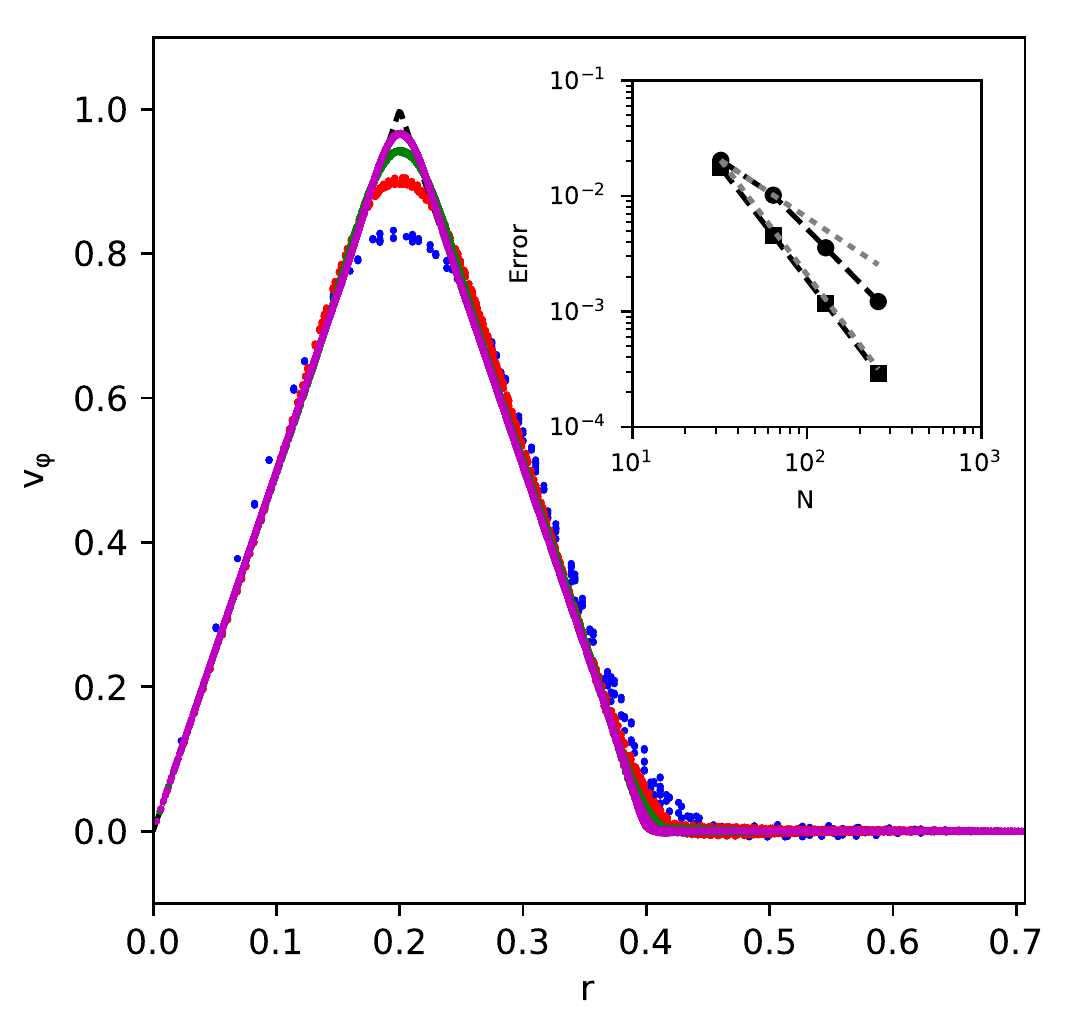}
\includegraphics[width=0.49\textwidth]{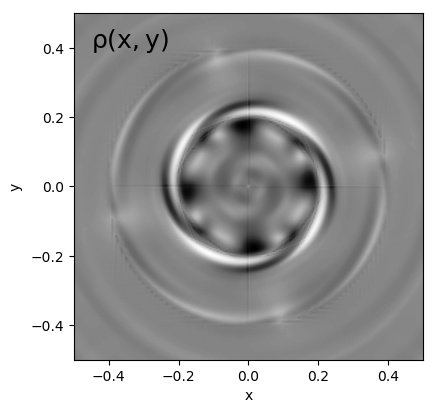}
}
\caption{The azimuthal velocity profile (left panel) for the Gresho-Chan vortex test, at increasing resolutions, in a scatter plot from the Cartesian 2D data. The inset quantifies the 1-norm for the rotation profile deviation (solid circles) and the 2-norm error in the pressure profile (squares) and we find better than 1st and up to second order convergence. The right panel shows the density distribution at time $t=2$. The variation in density is minute (order $1/10000$), but a physically meaningful pattern emerges, indicating the liability of this equilibrium to a linear instability. }\label{f-hdgresho}
\end{figure}

In the astrophysics literature, one finds many implementations of the HD system which exploit Smooth Particle Hydrodynamics (or SPH) techniques. These are meshless treatments of the governing conservation laws, and are in common use for large-scale cosmological simulations. A frequent test for any novel SPH variant, or for general new HD codes~\cite[e.g.][]{Chang2017,Wadsley2017,Mocz2014,Hopkins2015}, is the so-called Gresho-Chan vortex~\cite{Gresho1990}, which is believed to be a stationary solution to the Euler system where a pressure gradient balances the centrifugal force of a rotating gas. The initial condition generally has the form
\begin{eqnarray}
p & = & \left\{ \begin{matrix} \frac{1}{\gamma M^2} +12.5 r^2 &  &  0\leq r<0.2 \\
                      \frac{1}{\gamma M^2}+12.5 r^2 +4\left(1-5r-\ln(0.2)+\ln(r)\right)  & & 0.2<r\leq0.4 \\ 
                       \frac{1}{\gamma M^2}-2+4\ln(2)  & & r>0.4\end{matrix}\right.  \,, \label{gcp}
\end{eqnarray}
for the pressure, while the rotation velocity is simply
\begin{eqnarray}
v_\varphi & = & \left\{ \begin{matrix} 5r &  &  0\leq r<0.2 \\
                      2-5r  & & 0.2<r\leq0.4 \\ 
                      0  & & r>0.4\end{matrix}\right.  \,. \label{gcv}
\end{eqnarray}
A uniform density $\rho=1$ completes the setup, where the Mach number $M$ can be varied, a typical value is $M=0.34641=\sqrt{3}/{5}$. We use a square domain $[-0.5,0.5]\times [-0.5,0.5]$, translating the flow setup to $v_x$ and $v_y$ velocities, and use a one-step high-order TVD method with a Monotonized Central (or {\tt `woodward'}) limiter~\cite{Toth1996}. This scheme exploits the full approximate Roe solver that is aware of the characteristic decomposition in its flux computation. We solve up to $t=2$, with a Courant number of 0.9. This problem is solved on uniform meshes (no AMR), of size $32\times 32$, $64\times 64$, $128\times 128$ and $256\times 256$, respectively (our block size is always $8\times 8$) to demonstrate proper convergence. Info on the solution obtained is in Fig.~\ref{f-hdgresho}, where the left panel shows a scatter plot of the radial profile of $v_\varphi(r)$ (similar to the Cartesian version of our dam break problem in Fig.~\ref{f-sw}), this time for all the mesh sizes exploited. One can see that the azimuthal velocity nicely converges to the analytic initial condition, and the convergence behavior is quantified in the inset in the left panel, where the 1-norm of the error in the obtained azimuthal velocity profile is shown with circular symbols, while the 2-norm for the error in the pressure profile is shown with squares. To guide the eye, dotted lines indicating first as well as second order convergence are also provided. One observes that the pressure profile converges with second order accuracy, while the azimuthal velocity also behaves better than 1st order as soon as sufficient grid points are exploited (the $32\times 32$ run has roughly 12 grid points through the disk radius). Modern SPH variants demonstrate typically a $N^{-0.8}$ convergence rate in 1-norm~\cite{Wadsley2017}, where $N$ is then the equivalent 1D grid size, and hence only reveal sublinear convergence.

At right, the density profile at $t=2$ for the  highest resolution simulation is shown as well. To appreciate the scale, the density ranges between $[1.0001473,0.9998174]$.  The density variation obtained suggests that the stationary equilibrium configuration, which is typified by two specific radii where derivatives behave discontinuously, may well be subject to a higher $m$-mode instability, with a variation in azimuth angle $\propto m\varphi$, and indeed the longer term evolution eventually deviates from the initial setup. A rigorous stability analysis of the 1D rotating equilibrium confirms that the equilibrium is liable to a number of unstable linear eigenmodes, with e.g. a global overstable $m=2$ mode. It is pointed out here that many of the published results obtained with SPH variants report a lesser degree of convergence, and ignore the fact that the setup may be intrinsically unstable, casting doubt on quantifying errors at even later times than those used here.

\subsubsection{K\'{a}rm\'{a}n vortex streets}

A final HD test problem adds viscous source terms to the momentum and energy equations, where we intend to simulate viscous, time-dependent flow about a cylindrical obstacle. We coded up viscosity terms corresponding to
\begin{eqnarray}
\partial_t(\rho\bfv) & = & -\nabla\cdot(\nu \hat{\Pi}) \,, \\
\partial_t E & = & -\nabla\cdot(\bfv\cdot \nu \hat{\Pi}) \,,
\end{eqnarray}
where we introduce the traceless part of the kinetic pressure dyad through 
\begin{equation}
\hat{\Pi}= - \left((\nabla \bfv) +(\nabla\bfv)^T\right) +\frac{2}{3}\bfI (\nabla\cdot\bfv) \,.
\end{equation}
Note that we do treat these terms as sources, although their divergence-form would also allow one to include them in the flux definitions. As these source terms are identical for hydrodynamic and MHD applications, the viscous source terms are encoded in a module {\tt amrvac\//src\//physics\//mod\_viscosity.t}, which is then shared between the HD and MHD systems.

\begin{figure}[t]
\centerline{\includegraphics[width=0.8\textwidth]{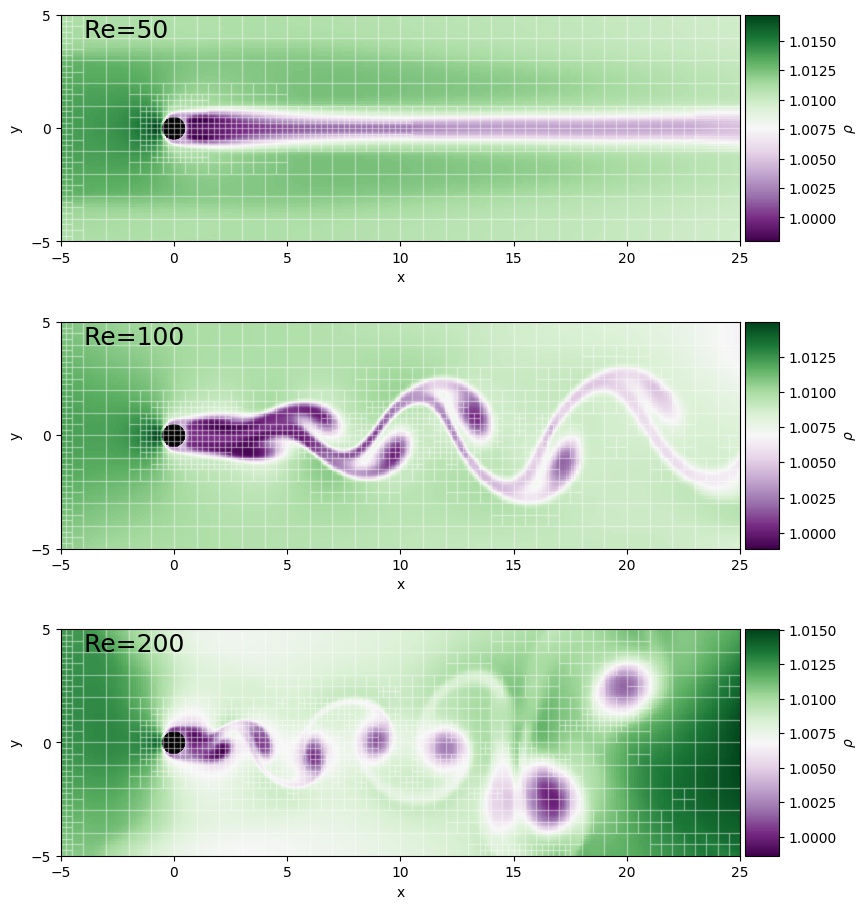}}
\caption{Compressible hydro simulations of flow about a cylinder, for fixed Mach number $M=0.1$, at varying Reynolds number $R_e=50$, 100, to 200. We show the density variation at $t=85$.}\label{f-hdkarman}
\end{figure}

For incompressible HD, where the velocity field is constrained by $\nabla\cdot\bfv=0$, it is known from experimental observations that only the Reynolds number $R_e$ is relevant in determining the flow properties downstream of the obstacle~\cite{Frisch}. For a typical flow speed $v_0$ and lenghtscale $l_0$, the Reynolds number sets the viscosity coefficient $\nu$ through $R_e=v_0 l_0/\nu$. As initial condition, we set units through $l_0=1$, $v_0=1$, and set up left-right symmetric potential flow about the cylinder, which has radius $r_0=l_0/2$. The detailed profiles are given by
\begin{eqnarray}
v_x & = &  1+\frac{r_0^2}{r^2}-\frac{2x^2r_0^2}{r^4} \,,\\
v_y & = & -\frac{2xyr_0^2}{r^4} \,,\\
p & = & p_0+ \frac{1}{2}\left( \frac{2 r_0^2\cos(2\theta)}{r^2}-\frac{r_0^4}{r^4}\right) \,,
\end{eqnarray}
where $r^2=x^2+y^2$ along with $\cos(\theta)=x/r$.
The density is uniform and $\rho=1$ initially, while $p_0=1/\gamma M^2$, introducing a Mach number $M$ when simulating this setup with the compressible HD system. To remain close to the expected behavior for incompressible flow, we set $M=0.1$. We simulated three cases, which only differ in the Reynolds number $R_e$, varied from 50, 100, to 200. The left inlet boundary exploits Dirichlet boundary conditions, setting a uniform horizontal inflow corresponding to the far-field solution where $\rho=1$, $p=p_0$ and $v_x=1$. The other three lateral boundaries use a zero gradient extrapolation. Special to this setup is an approximate treatment of the internal region $r<0.5$ for the cylinder: we actually nullify the full flow field within this radius, mimicking the vanishing of the flow components (in a viscous boundary layer) expected for a Navier-Stokes evolution. We further use a domain $[-5,25]\times[-5,5]$, employ a base level grid of $300\times 100$, and use AMR allowing 3 refinement levels in total. We enforce full refinement manually about the cylinder, and let the remainder of the domain regrid on the basis of variations in density and horizontal momentum $\rho v_x$. 

A strong stability preserving Runge-Kutta scheme~\cite{mpiamrvac2014}, combined with an HLL flux~\cite{hll} and Koren reconstruction~\cite{Koren1993} is used, and we show the obtained solutions at $t=85$ in Fig.~\ref{f-hdkarman}. Note that these figures show the density variation, different from the incompressible situation where the density is uniform throughout. However, we do recover the correct transition to turbulent flows, as the Reynolds number increases, which shows how gradual symmetry breaking occurs for higher Reynolds numbers. The $R_e=50$ solution settles on a steady state, where the original left-right symmetry of the potential flow is broken, but the up-down symmetry preserved. Higher Reynolds numbers break the up-down symmetry spontaneously, first showing fairly regular, periodic vortex shedding ($R_e=100$), while even this transits to more chaotic behavior as $R_e$ reaches 200. This is the typical behavior of K\'{a}rm\'{a}n vortex streets~\cite{Frisch}. Our compressible simulations also show some sound wave related background variations in density induced by artificial reflections (at inlet and at the cylinder). 

\begin{figure}[ht]
\centerline{\includegraphics[width=0.8\textwidth]{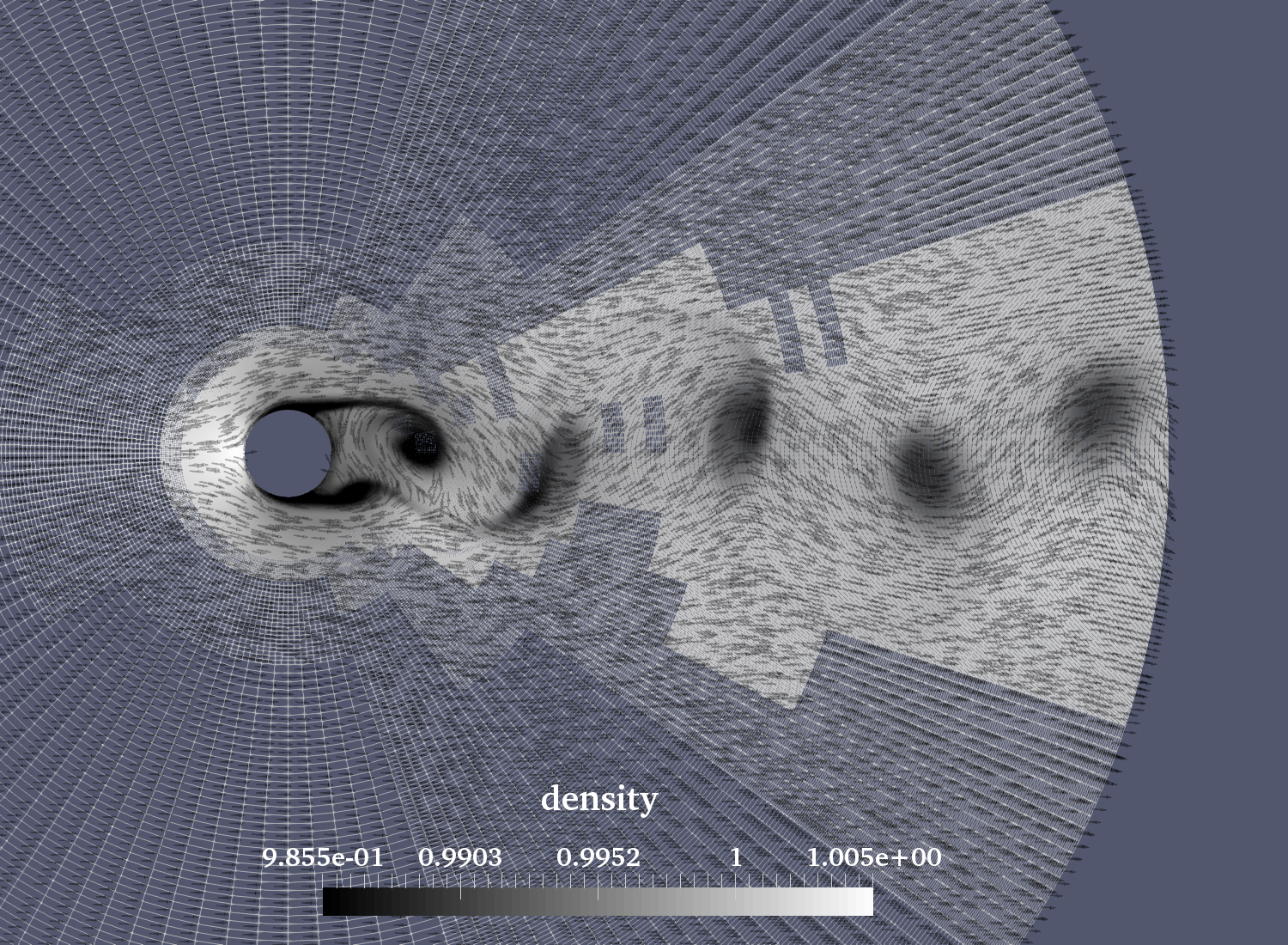}}
\caption{Compressible flow about a cylinder, computed on a polar AMR mesh, for Mach $M=0.1$, at Reynolds number $R_e=200$. We show the density variation at $t=85$ in greyscale, and the flow field using arrows, along with the grid structure.}\label{f-hdkarman2}
\end{figure}

The same setup can also be simulated on a 2D polar AMR mesh, where the boundary conditions for vanishing flow at the cylinder radius can be enforced exactly. An impression of a $M=0.1$, $R_e=200$ simulation is given in Fig.~\ref{f-hdkarman2}. In this setup, the outer radial boundary treatment is less trivial (potential inflow is enforced, while open flow conditions at the right half prevail). The pressure distribution as a function of polar angle along the cylinder radius can be compared to actual flow measurements, as this setup has been studied extensively in terms of the drag coefficient. The variation of the flow properties for increasing Reynolds number can then be verified, studying e.g. vortex shedding frequencies (i.e. Strouhal number). The use of AMR helps to affordably achieve a high resolution, capturing all details in the boundary layer (such as the separation angle, possible transitions to turbulence) and the wake region.

\subsection{MHD blast wave}

As an example MHD run, we use a frequently quoted MHD blast wave configuration~\cite[e.g.][]{Balsara2015,Felker2018,Yang2018,Gardiner2008}, which is in spirit similar to the circular dam break setup, but where the initially uniform magnetic field $\bfB$ now introduces a clear anisotropy. We use the exact setup recently demonstrated in 2D in~\cite{Felker2018}, where $\rho=1$, $\gamma=5/3$, the domain is $[-0.5.0.5]^2$ and the magnetic field is $\bfB=(1/\sqrt{2},1/\sqrt{2})$. A central circular region of radius $r_{\mathrm{blast}}=0.1$ has an overpressure $p_{\mathrm{in}}=10$, in contrast to the exterior $p_{\mathrm{ext}}=0.1$. The plasma beta, quantifying the dimensionless ratio $\beta=2p/B^2$ (exploiting units where the vacuum permeability $\mu_0=1$) ranges from 0.2 (outside) to 20 (inside the blast). The simulation is run till time $t=0.2$. 

\begin{figure}[t]
\centerline{\includegraphics[width=\textwidth]{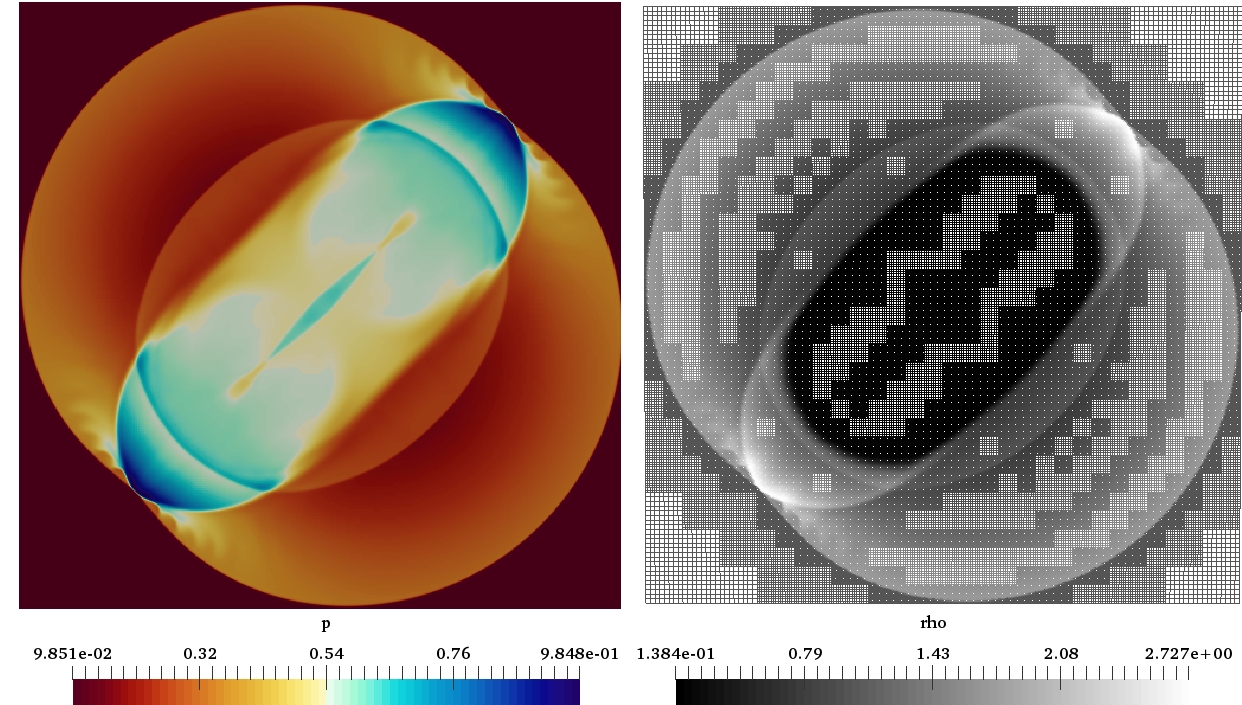}}
\caption{An MHD blast wave solution in 2D. The pressure (left) and density (right) are shown at $t=0.2$, and an impression of the adaptive grid structure is given in the right panel.}\label{f-2dmhd}
\end{figure}

As there is no real exact solution known, we simply show plots in a format that allows direct comparison with published 2D results. We use this setup to showcase the dimension and coordinate flexibility of our software, hence we will run it in 2D (on $[-0.5.0.5]^2$) and 3D Cartesian (then on $[-0.5.0.5]^3$) setups, as well as on a 2D polar and a 3D spherical grid. In~\cite{Yang2018}, similar 2D and 3D results for (nearly) identical setups were shown, on both Cartesian versus polar (2D) and spherical grids (3D). \cite{Yang2018} demonstrated the adaptation of a modern space-time conservation element and solution element (CESE) scheme on otherwise fixed, but on general curvilinear grids. This CESE scheme was also demonstrated with AMR and general curvilinear grids on MHD blast waves in~\cite{Jiang2010}. In 2D polar setups, we use $r\in[1,2]$ and $\varphi\in[-0.12\pi,0.12\pi]$, with the initial blast at $r=1.5$, $\varphi=0$, as in~\cite{Yang2018}. Similarly, for the 3D Cartesian run, we set the initial field $\bfB=(1/\sqrt{3},1/\sqrt{3},1/\sqrt{3})$ to retrieve the same plasma beta regime. In the 3D spherical run, we will simulate on a shell $r\in[0.1,1.1]$, while polar angle $\vartheta\in[0.2\pi,0.8\pi]$ and angular variation $\varphi\in[0.7\pi,1.3\pi]$ is used. In this spherical setup, we initialize $\bfB=(1/\sqrt{2},1/\sqrt{2},0)$, while putting the blast perturbation at $r_b=0.6$, $\vartheta_b=\pi/2$ and $\varphi_b=\pi$. We will use AMR in all runs.

\begin{figure}[t]
\centerline{\includegraphics[width=\textwidth]{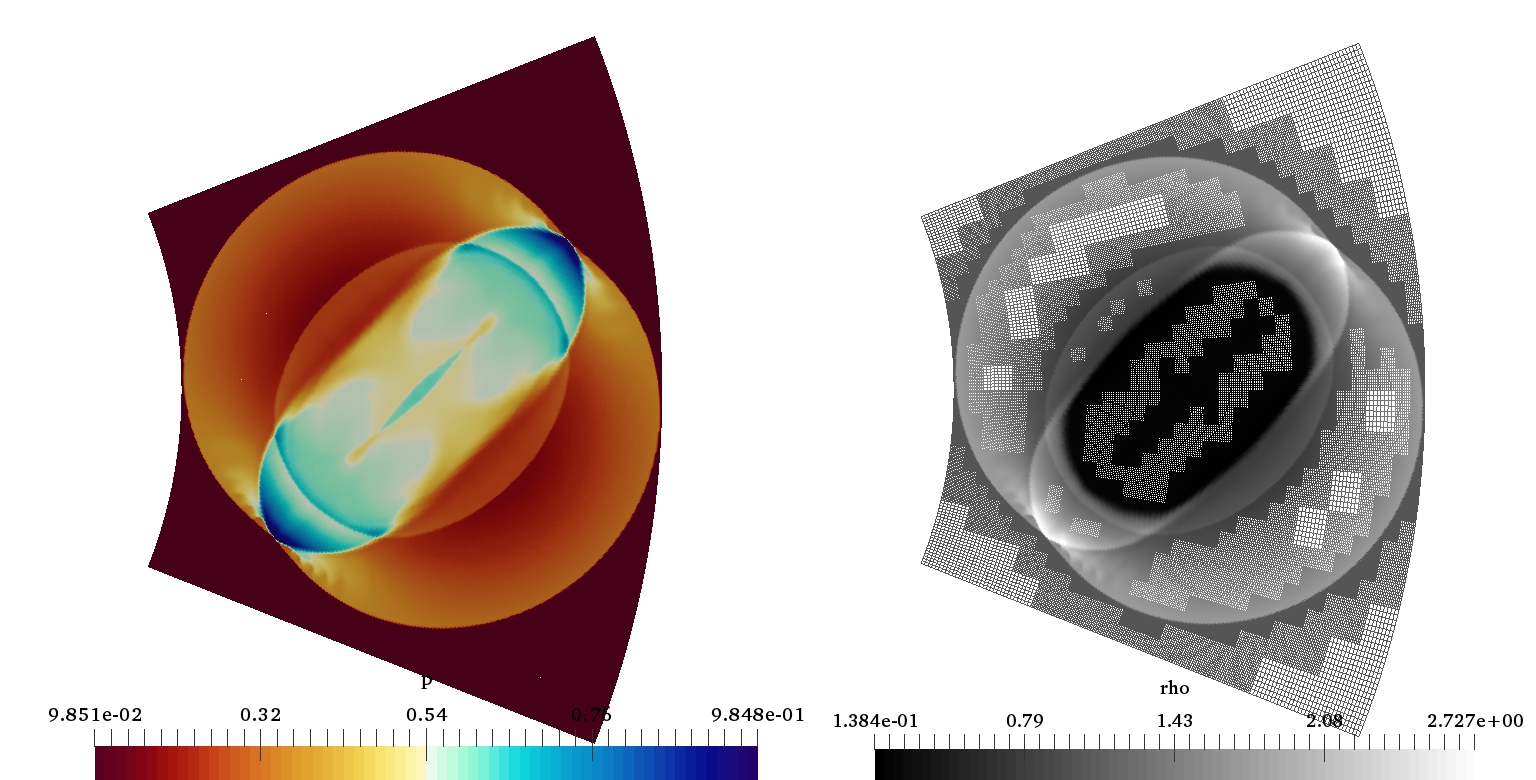}}
\caption{An MHD blast wave solution in 2D, simulated on a polar AMR grid. In the same format as Fig.~\ref{f-2dmhd}, the pressure (left) and density (right) are shown at $t=0.2$, and an impression of the adaptive grid structure is given in the right panel.}\label{f-2dmhdb}
\end{figure}

We always use a three-step time integration, an HLLC flux computation~\cite{toro97,Li2005}, and the third order Cada limiter~\cite{cada} in the reconstruction. A Courant parameter of 0.9 is used, but the first 10 discrete timesteps are reduced gradually. At $t=0.2$, the perturbation has not yet reached any lateral boundary, so the boundary conditions are fairly irrelevant: we used periodic sides in the Cartesian setups, and fixed all quantities in the polar and spherical case. The 2D runs have base resolution $64 \times 64$, with 3 additional refinement levels hence reaching $512^2$ thanks to the AMR, while the 3D runs use $64^3$, up to effective resolutions $256^3$ thanks to the AMR. 

\begin{figure}[th]
\centerline{\includegraphics[width=0.48\textwidth]{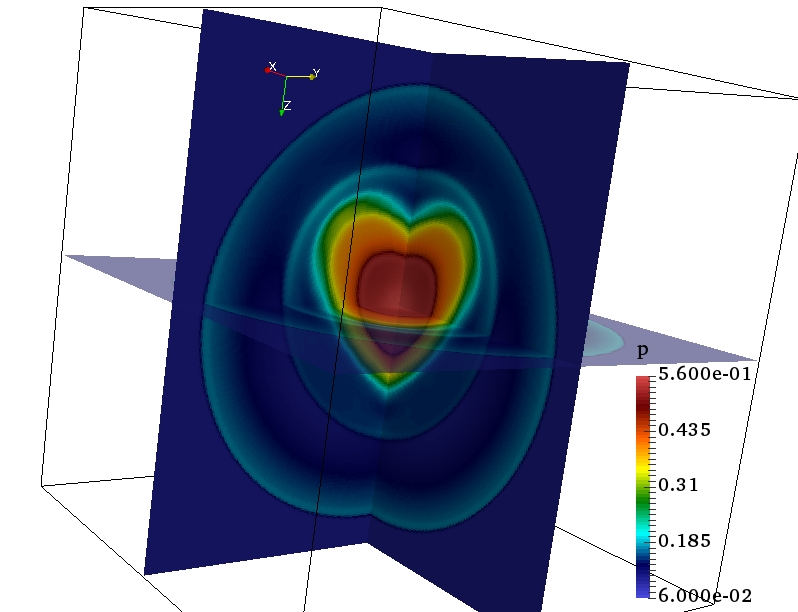}
\includegraphics[width=0.48\textwidth]{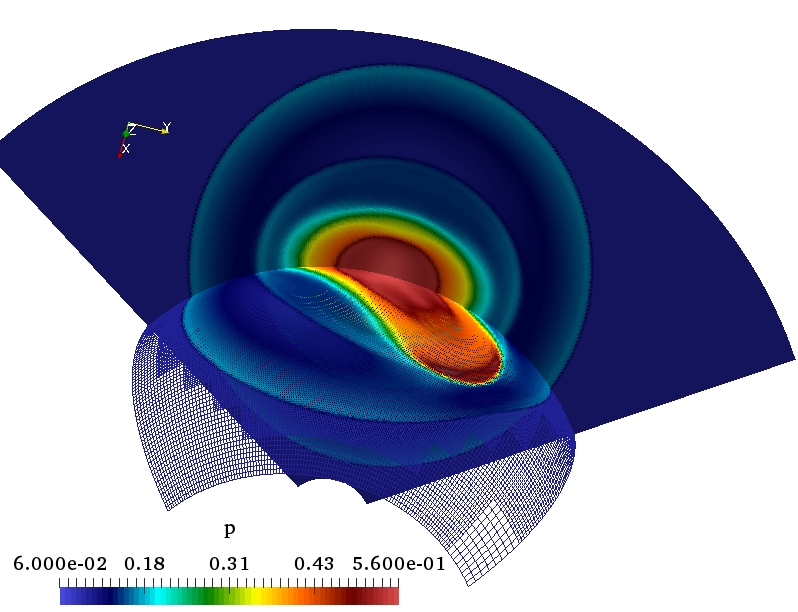}}
\caption{An MHD blast wave solution in 3D. On a Cartesian grid (left), showing the pressure on three orthogonal cutting planes (with the $z=0$ one made translucent). On a block-AMR spherical grid (right).}\label{f-3dmhd}
\end{figure}

In Fig.~\ref{f-2dmhd}, the pressure (left) and density (right) are shown in contour views, where the grid structure is visible in the density panel. In this setup, the initial magnetic field is oriented along the diagonal, and the various MHD wave signals cause intricate patterns. Note in particular the pressure-density fluctuations in the north-east and south-west disturbed regions in between the outermost (fast) shock front and the more elliptical shaped signal (outlined by the blueish color in the pressure view). When repeating the simulation on a polar grid, shown in Fig.~\ref{f-2dmhdb}, the same details emerge, although in that case, the south-west part is slightly more resolved than the north-east perturbation, due to the natural $r$-dependence in the polar grid structure. Our Fig.~\ref{f-2dmhd} compares favorably with the results shown in~\cite{Felker2018}, who computed on a fixed grid, using a novel fourth-order finite volume method using a constrained transport approach for handling $\nabla\cdot\bfB=0$. Our AMR run from Fig.~\ref{f-2dmhdb} shows more details than the polar result in~\cite{Yang2018}, due to its higher effective resolution.

We emphasize that our simulation does not exploit any staggering (all quantities are defined cell-centered), and on both the 2D and the 3D Cartesian grid, we used the multigrid functionality to control $\nabla\cdot\bfB=0$ as described in~\cite{mg2019}. This in practice implies the multigrid based solution of a Poisson problem $\nabla^2 \phi = \nabla\cdot\bfB^*$ where $\bfB^*$ is the magnetic field after a (sub)step of any scheme applied, to correct it to a solenoidal $\bfB=\bfB^*-\nabla\phi$. As our multigrid solver can not handle the grid variation from a polar or spherical mesh, the way to control magnetic monopole errors in those runs was using the diffusive treatment introduced in~\cite{Keppens2003}, only applied to the induction equation.

The 3D simulation on the Cartesian grid is shown at left in Fig.~\ref{f-3dmhd}, where we show the instantaneous pressure distribution on 3 cutting planes ($x=0$, $y=0$ and $z=0$). Note that there are (expected) notable differences between the purely 2D and the 3D blast evolution, as recovering the 2D result would require a cylindrical, instead of a spherical, initial blast region. The final blast wave demonstration is the spherical 3D simulation, shown at right in Fig.~\ref{f-3dmhd}.  We again show the pressure distribution, shown on two surfaces ($r=0.6$ which also gives an impression of the mesh, and $\vartheta=\pi/2$). The results compare favorably with similar 3D tests in~\cite{Yang2018}.

\subsection{Reaction-Diffusion models}
\label{sec:react-diff}

Although the last \texttt{A} in \texttt{MPI-AMRVAC} stands for \emph{advection},
the code can also handle problems without advection. The reaction-diffusion
{\tt amrvac\//src\//rd} module, which has recently been added to \texttt{MPI-AMRVAC}, can be used to
solve equations with two chemical concentrations\footnote{It can trivially
  be extended to more than two chemical species.}.
Such systems can exhibit a wide variety of pattern-forming
behavior~\cite{Kondo_2010}, as was first pointed out by
Turing~\cite{Turing_1990}. We specifically consider two types of models, the
first being the Gray-Scott model~\cite{Gray_1983}, which in dimensionless units
has the following form
\begin{align}
  \label{eq:gray-scott}
  \partial_t u &= D_u \nabla^2 u - u v^2 + F(1 - u),\\
  \partial_t v &= D_v \nabla^2 v + u v^2 - (F + k) v\nonumber,
\end{align}
where $F$ and $k$ are positive constants, and the diffusion coefficients are
here set to $D_u = 2 \times 10^{-5}$ and $D_v = 10^{-5}$. Note that $u$ is
converted to $v$, and that the `feed' term $F(1 - u)$ drives the concentration
of $u$ to one, whereas the term $-(F + k) v$ removes $v$ from the system.
Depending on the values of $F$ and $k$, a wide range of patterns can be
generated, as demonstrated in~\cite{Pearson_1993}. Here we use $F = 0.046$ and
$k = 0.063$.

The second type of model we
consider is due to Schnakenberg~\cite{Schnakenberg_1979}
\begin{align}
  \label{eq:schnakenberg}
  \partial_t u &= D_u \nabla^2 u + \kappa (a - u + u^2 v),\\
  \partial_t v &= D_v \nabla^2 v + \kappa (b - u^2 v),
\end{align}
where $\kappa$, $a$ and $b$ are positive constants. The reaction terms somewhat
differ from the Gray-Scott model, but the most important difference is that we
will use much larger diffusion coefficients: $D_u = 0.05$, $D_v = 1$. These and
other parameters ($\kappa = 100$, $a = 0.1305$ and $b = 0.7695$) are taken
from~\cite{Hundsdorfer_2013} (Chapter IV, section 4.4).

\subsubsection{Numerical implementation}
\label{sec:implementation}

The implementation of these reaction-diffusion models in
\texttt{MPI-AMRVAC} is handled via source terms, using the {\tt
  phys\_add\_source} interface. A standard second-order accurate discretization
of the diffusive terms is employed, and fluxes are not considered in this
module. The reaction terms are always handled explicitly, but for the diffusion
terms we have implemented several options. The first is to handle diffusion
explicitly, which leads to a time step restriction
$\Delta t < h^2 / (2 N_\mathrm{dim} D_\mathrm{max})$, where $h$ is the grid spacing,
$N_\mathrm{dim}$ is the problem dimension and
$D_\mathrm{max} = \max(D_u, D_v)$ the maximum of the diffusion coefficients.
Explicit time step restrictions for the reaction and diffusion terms are
tabulated in Table~\ref{tab:dt-diffusion}. The large diffusion coefficients make
the Schnakenberg model numerically stiff, even on relatively coarse grids.

\begin{table}
  \centering
  \begin{tabular}{c|c|ccc}
    model & reactions & $h = 1/128$ & $1/256$ & $1/512$\\
    \hline
    Gray-Scott & $\sim 1$ & $0.76$ & $0.19$ & $0.048$\\
    Schnakenberg & $\sim 10^{-2}$ & $1.5 \times 10^{-5}$ & $3.8 \times 10^{-6}$ & $0.95 \times 10^{-6}$
  \end{tabular}
  \caption{Restriction on $\Delta t$ due to reactions (first column) and due to
    handling diffusion explicitly for a 2D problem with grid spacing $h$.}
  \label{tab:dt-diffusion}
  % Reaction terms: $\Delta t < 1/(v^2 + F)$ and $\Delta t < 1/|uv - F - k|$
\end{table}

A detailed comparison of numerical methods to handle stiff reaction-diffusion
problems can be found in \cite{Hundsdorfer_2013,Ruuth_1995}. In
\texttt{MPI-AMRVAC}, we have implemented two schemes. The first is a simple
operator splitting method. The idea is to split the time derivative as
\begin{equation}
  \label{eq:splitting}
  \partial_t w = F(w) = F_0(w) + F_1(w),
\end{equation}
where $F_0$ are the non-stiff reaction
terms and $F_1$ the stiff diffusion terms. The effect of $F_0$ can be handled
explicitly to obtain $w_{n+1}^*$ from a past state $w_n$, after which an
implicit equation is solved to obtain the next state $w_{n+1}$. We use a
backward-Euler discretization $w_{n+1} = w_{n+1}^* + \Delta t \, F_1(w_{n+1})$,
which leads to a Helmholtz equation:
\begin{equation}
  \nabla^2 w_{n+1} -\frac{1}{\Delta t D} w_{n+1} = -\frac{1}{\Delta t D} w_{n+1}^*.
  \label{eq:diff-helmholtz}
\end{equation}
This equation is solved with the parallel and AMR-compatible geometric multigrid
solver that has recently been added to
\texttt{MPI-AMRVAC}~\cite{mg2019}. Such a multigrid method leads to a
linear cost in the number of unknowns.

\begin{figure}[t]
  \centering
  \includegraphics[width=\textwidth]{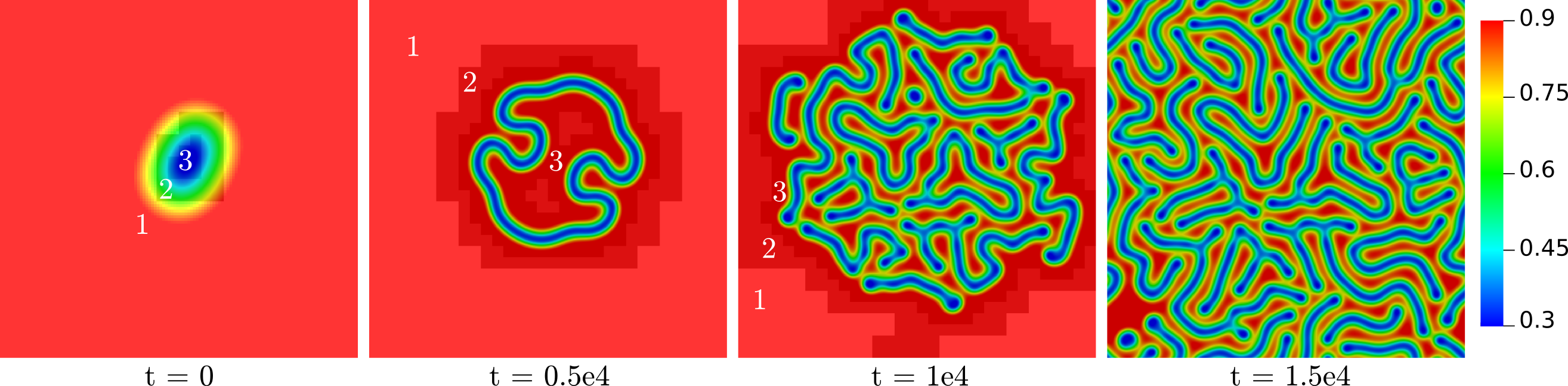}
  \caption{Time evolution of the density $u$ in a Gray-Scott model with
    $F = 0.046$ and $k = 0.063$, see equation \eqref{eq:gray-scott}. Three refinement levels are used, indicated by the white number and the
      gray shade. The levels correspond to an effective resolution of $128^2$,
      $256^2$ and $512^2$ cells. In the rightmost picture, the whole grid is at
      the highest refinement level.}
  \label{fig:gray-scott}
\end{figure}

The second scheme we have implemented is the second-order accurate IMEX scheme given in
\cite{Hundsdorfer_2013} (eq.~4.12 of chapter IV), which is a combination of the
implicit and explicit trapezoidal rule:
\begin{align}
  w_{n+1}^* &= w_n + \Delta t F_0(w_n) + \tfrac{1}{2} \Delta t \left [F_1(w_n) + F_1(w_{n+1}^*)\right]\\
  w_{n+1} &= w_n + \tfrac{1}{2} \Delta t \left[F(w_n) + F(w_{n+1}^*)\right]. \label{imex}
\end{align}
When written out, the first equation again corresponds to a Helmholtz equation
that can be solved with the multigrid solver.

\begin{figure}[t]
  \centering
  \includegraphics[width=\textwidth]{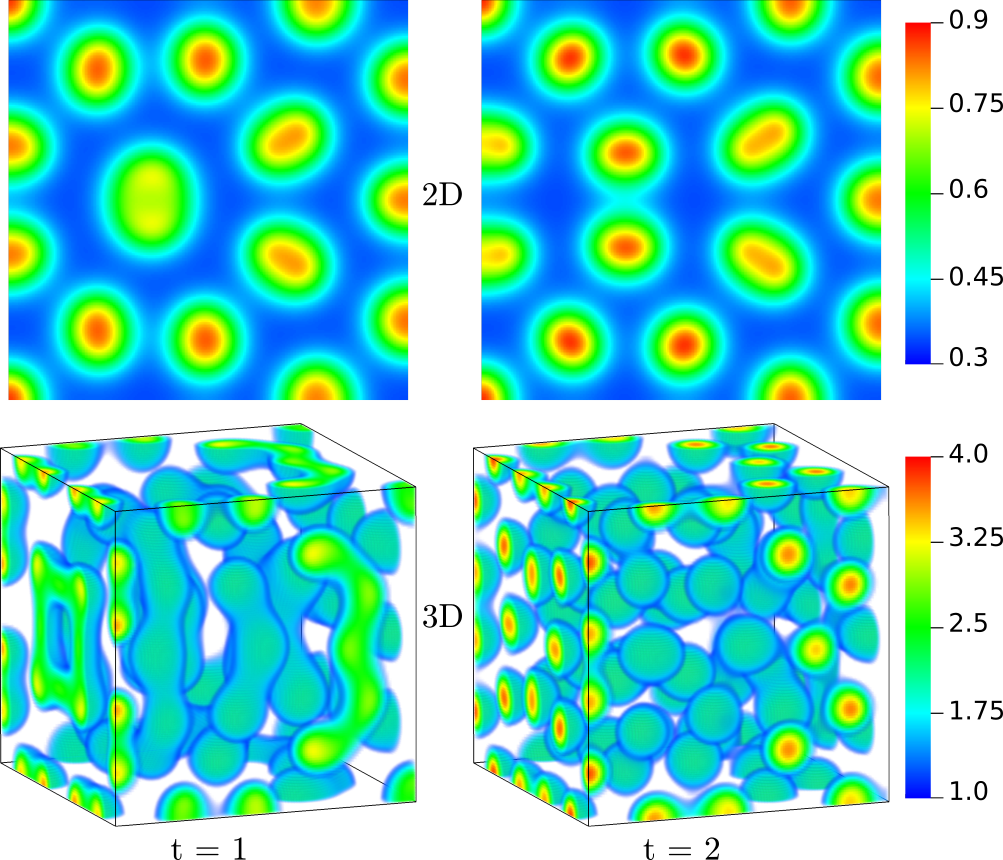}
  \caption{Time evolution of the density $u$ in Schnakenberg's model, see
    equation \eqref{eq:schnakenberg}. Top row: 2D case on a $256^2$
      uniform grid, bottom row: 3D case on a $128^3$ uniform grid.}
  \label{fig:schnakenberg}
\end{figure}

\subsubsection{Examples}
\label{sec:react-diff-examples}

Figure \ref{fig:gray-scott} shows the time evolution of a Gray-Scott model for
which $F = 0.046$ and $k = 0.063$. The model is solved up to
$t = 1.5 \times 10^4$ in a periodic 2D domain of size $L \times L$, with
$L = 2$.
An AMR mesh with three levels is used, corresponding to grids of $128^2$ up to $512^2$ cells, and the size of individual grid blocks is set to $8^2$ cells.
Time integration is performed with
the midpoint method using a time step $\Delta t = 0.5$. The initial condition is
the steady state $u = 1$ and $v = 0$ modified by two Gaussian perturbations of
the form $\tfrac{1}{2} \exp(-25 |\vec{r} - \vec{r}_i|^2)$, with
$\vec{r}_1 = (0.5, 0.5)$ and $\vec{r}_2 = (0.55, 0.6)$. These perturbations are
subtracted from $u$ and added to $v$. A complex maze-like pattern emerges. For
other values of $F$ and $k$, many other types of patterns can emerge, see
\cite{Pearson_1993}\footnote{Interested readers can also interactively explore
  such patterns at \url{https://mrob.com/pub/comp/xmorphia/ogl/index.html}}.

\begin{figure}[t]
  \centering
  \includegraphics[width=0.8\textwidth]{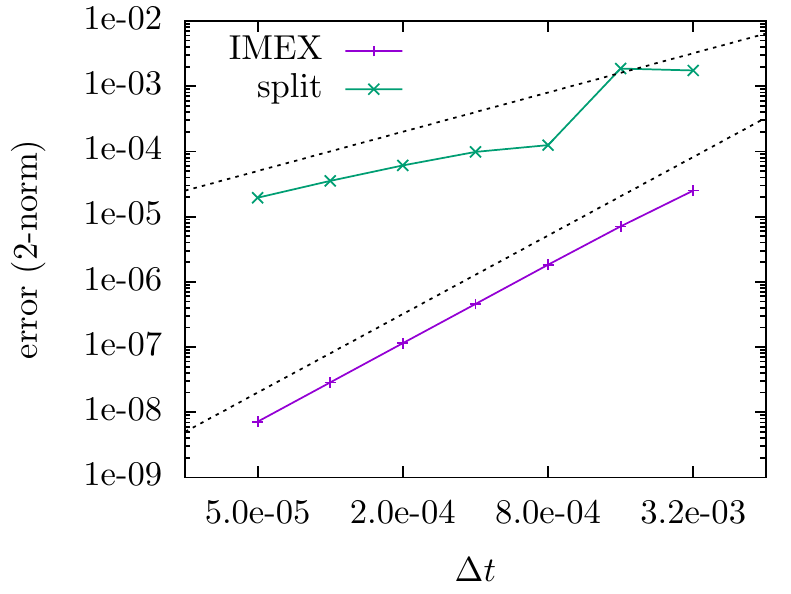}
  \caption{Time integration error (two-norm) of an IMEX and a split scheme for
    solving Schnakenberg's model on a $256^2$ uniform grid.
    The solution at $t = 2$ is compared to a
    solution computed with a small time step. The dashed lines indicate first
    and second order convergence.}
  \label{fig:schnakenberg-conv}
\end{figure}

The evolution in figure \ref{fig:gray-scott} is somewhat chaotic and therefore
sensitive to small numerical errors. To compare the numerical properties of
reaction-diffusion schemes, we consider a 2D and 3D example in which we solve
Schnakenberg's model, which has a less chaotic time evolution. Solution examples
are shown in figure \ref{fig:schnakenberg}, both for 2D and 3D cases. As in
\cite{Hundsdorfer_2013}, we use a domain with sides of length $L=1$, and Neumann
zero boundary conditions for the species densities. The initial condition is
$v = b/(a+b)^2$ and $u(\vec{r}) = a + b + \exp(-100 |\vec{r} - \vec{r}_0|^2)$,
where $\vec{r}_0 = (1/3, 1/2)$ in 2D and $\vec{r}_0 = (1/3, 1/2, 1/2)$ in 3D.

Figure \ref{fig:schnakenberg-conv} shows the convergence behavior of the IMEX
and the split scheme for Schnakenberg's problem in 2D solved on a uniform grid of
$256^2$ cells. The solution at $t = 2$ is compared to a solution computed with
an explicit third-order scheme and a small time step
$\Delta t = 3\times 10^{-6}$. The IMEX scheme (from Eq.~(\ref{imex})) performs well and exhibits second
order convergence. The split scheme converges more slowly, with slightly less
than first order convergence, which indicates that there are large splitting
errors. These results are in agreement with~\cite{Hundsdorfer_2013}. 

To compare the computational costs of the schemes in \texttt{MPI-AMRVAC}, we ran
the $256^2$ test case using 4 cores of an AMD 2700X CPU. Per time step, each
scheme took: explicit 1.1 ms, IMEX 7.7 ms and split 7.2 ms. For the latter two
schemes, the multigrid solver consumed about 90\% of the CPU time. This
percentage is so high because the reaction terms are computationally cheap to
evaluate. For the multigrid solver, iterations were performed until the maximum
residual was less than $10^{-7}$ times the right-hand side of equation
\eqref{eq:diff-helmholtz}. The explicit scheme is the cheapest per time step,
but it requires orders of magnitude more steps than the other methods for
stability.

\section{Conclusions and outlook}

We gave an overview of currently available PDE systems in the open-source software {\tt MPI-AMRVAC}, demonstrating its versatility in dimensionality, but also in the type of PDE systems to be solved. For the (M)HD system, to which it was originally targeted, various conservative, shock-capturing discretizations are implemented. With minimal effort, any near-conservative system may be implemented as a new physics module, and the framework offers a dimension-independent, parallelized means for performing high-resolution, domain decomposed or block grid-adaptive computations. The recent coupling to a geometric multigrid solver extends its versatility to any problem where Poisson or Hemholtz type constraints arise, and this was demonstrated here for a newly added reaction-diffusion set of equations. For stiff source terms, such as particular diffusion terms in these reaction-diffusion problems, we can use our framework to compare modern variants of IMEX schemes, with standard explicit treatments. We documented here how a new PDE system is readily added to the source code, and welcome any extension of our software to explore intricate spatio-temporal behavior of nonlinear PDEs.

As illustrated on both shallow water and MHD Riemann problems, the code can handle simulations on polar, cylindrical or spherical grids, which require the handling of geometric source terms. Generic gradient, vector divergence and vector curl operations are implemented in the {\tt amrvac\//src\//mod\_geometry.t} module, and they can all be combined with directional stretching (e.g. demonstrated for radial directions in~\cite{mpiamrvac2018}). As stated earlier, {\tt MPI-AMRVAC} has been used succesfully to handle not only Newtonian (M)HD, but its extension to special relativistic (M)HD as well~\cite{mpiamrvac2012}, where also a typical 3+1 space-time formulation leads to a system of the form~(\ref{eq}). Advanced applications to the extreme conditions encountered in pulsar wind nebulae~\cite{crab1,crab2} focused on such relativistic plasma behavior in Minkowski space-time. Meanwhile, code variants that can handle also non-orthogonal curvilinear coordinates, where one must distinguish between covariant and contravariant vector representations, have been developed~\cite{gramrvac2016,bhac2017}. The {\tt Black Hole Accretion Code} or {\tt BHAC}~\cite{bhac2017} solves the covariant general relativistic MHD (GRMHD) equations in a 3+1 foliation of space time, where a flexible data structure has been introduced to handle any four-metric. A recent code comparison project~\cite{Porth2019} between the most modern software efforts to simulate GRMHD conditions as suitable in the vicinity of black holes showed that {\tt BHAC} meets all standards of merit for guiding and interpreting contemporary astrophysical research. The {\tt BHAC} code has recently been extended with an IMEX scheme to handle the extension to general relativistic, resistive MHD equations~\cite{grrmhd2019}, where all covariant Maxwell equations, especially also those for electric field evolutions, enter. The IMEX scheme then treats the stiff resistive source terms, and can use a staggered representation of the GR(R)MHD variables, to ensure that magnetic monopoles only occur at machine precision~\cite{Olivares}. The code structure and modularity, especially in its parallelization and AMR strategy, is fully shared between the {\tt MPI-AMRVAC} and {\tt BHAC} code efforts. 

Future work can extend the code applicability to incompressible (M)HD regimes, kinematic dynamo studies where only the induction equation for the evolution of $\bfB$ from MHD is handled, or applications involving self-gravity. 

\section*{Acknowledgements}
RK was supported by a joint FWO-NSFC grant G0E9619N and by his ERC Advanced Grant PROMINENT, and JT was supported by FWO postdoctoral fellowship 12Q6117N.
This project has received funding from the European Research Council (ERC) under the European Union’s Horizon 2020 research and innovation programme (grant agreement No. 833251 PROMINENT ERC-ADG 2018). This research is further supported by Internal funds KU Leuven, project $C14/19/089$ TRACESpace. The
computational resources and services used in this work were provided by
the VSC (Flemish Supercomputer Center), funded by the Research Foundation
Flanders (FWO) and the Flemish Government - department EWI.  

\section*{References}

\bibliography{CAMWA2019}

\end{document}